# CORE EXCITED FANO-RESONANCES IN EXOTIC NUCLEI

S.E.A.Orrigo[a,b,*], H.Lenske[c], F.Cappuzzello[a], A.Cunsolo[a,b],

A.Foti[b,d], A.Lazzaro[a], C.Nociforo[a], J.S.Winfield[a]

a) INFN Laboratori Nazionali del Sud, Catania, Italy

b) Dipartimento di Fisica e Astronomia, Università di Catania, Catania, Italy

c) Institut für Theoretische Physik, Universität Giessen, Giessen, Germany

d) INFN Sezione di Catania, Catania, Italy

**Abstract.** Fano-resonances are investigated as a new continuum excitation mode in exotic nuclei. By theoretical model calculations we show that the coupling of a single particle elastic channel to closed core-excited channels leads to sharp resonances in the low-energy continuum. A signature for such Bound States Embedded in the Continuum (BSEC) are characteristic interference effects leading to asymmetric line shapes. Following the Quasiparticle-Core Coupling model we consider the coupling of 1-QP (one-quasiparticle) and 3-QP components and find a number of long-living resonance structures close to the particle threshold. Results for $^{15}$C are compared with experimental data, showing that the experimentally observed spectral distribution and the interference pattern are in qualitative agreement with a BSEC interpretation.

[*]Corresponding author: S.E.A. Orrigo, Via Santa Sofia 62, 95123 Catania, Italia, Tel. 0039(0)95542384, Fax. 0039(0)957141815, e-mail: orrigo@lns.infn.it.

## 1. Introduction

A new frontier of modern nuclear physics is the study of nuclei far off the valley of β-stability. Over the last few years, various unexpected phenomena have been observed, e.g. [1-6], indicating a shift away from mean-field dynamics towards a new type of correlation dynamics. Here, we consider a particular class of states above the neutron emission threshold, given by the coupling of single particle continuum states to closed channels involving core-excited configurations. Such interactions give rise to sharp resonance structures appearing in the low-energy continuum of the [n + (A − 1)] system. These states are known as Bound States Embedded in the Continuum (BSEC), which, for example, we have observed in $^{11}$Be [7] and $^{15}$C [8, 9]. Such Fano-resonances were originally detected in atomic spectra [10] and, since then, it has been realized that they are typical for interacting many-body systems at all scales.

The first model for auto-ionizing atomic states was developed by Fano in the 1960's [10]. Fano interference consists of the quantum-mechanical interaction between discrete and continuous configurations, leading to characteristically asymmetric peaks in the spectra. In particular, Fano found that the interaction of a discrete auto-ionized state with a continuum leads to an asymmetric line shape in the spectra of the emitted electrons [10]. The general phenomenon of the Fano interference has been observed in various branches of physics. Applications to atomic physics phenomena are reported, e.g., in ref. [11] (and further refs. therein). Recently, the method was applied in hadron physics for the investigation of the width and line shape of the ρ-meson in terms of the two-pion continuum [12]. The BSEC phenomenon in nuclei was predicted theoretically by Mahaux and Weidenmüller [13]. In ref. [14] it was investigated theoretically in stable nuclei and observed for the first time in the inelastic excitation of the $^{13}$C($3/2^+$) state at $E_x$ = 7.677 MeV [15]. In the case of inelastically scattered nucleons on nuclei, typical interference patterns were found where the energy dependence results in functions more complicated than simple Breit-Wigner distributions [14].

Since the BSEC appear as narrow resonances – often accompanied by an asymmetric line shape – at energies well beyond the neutron emission threshold, they cannot arise from simple resonances in a static potential. A natural explanation is that these structures are quasi-bound core-excited configurations, as described, e.g., by the Dynamical Core Polarization (DCP) model [16]. Close to the neutron drip line, the core itself is already neutron-rich. Thus an enhancement of BSEC structures is expected because of the softer core and the increased polarizability of the system. Also, the weak binding of the valence particles gives such nuclei the characteristics of an open quantum system. An interesting case is $^{15}$C which, with its small separation energy of the valence neutron ($S_n$ = 1.218 MeV), is intermediate between the well-bound $^{12,13,14}$C nuclei and the more exotic $^{17}$C and $^{19}$C ones.

$^{15}$C has been recently studied via the ($^7$Li,$^7$Be) Charge EXchange (CEX) reaction at the IPN-Orsay Tandem laboratory [8, 9]. Three narrow resonances were observed in the continuum at $E_x$ = (6.77 ± 0.06, 7.30 ± 0.06, 8.50 ± 0.06) MeV. Immediately before the peak at 8.5 MeV, a suppression of counts

is evident. This suppression, recognized as a dip, is clearly seen in spectra at various angles. In analogy to (p,p') inelastic scattering, where isolated resonances arising from the coupling of a single particle continuum to a BSEC have been found [14], the data suggest that this effect could be due to an interference between the non-resonant ($^{7}$Be + $^{14}$C + n) 3-body phase space, especially the $^{14}$C + n elastic 2-body background, and the 8.5 MeV BSEC. In refs. [8, 9] the 8.5 MeV peak was preliminarily fitted by assuming a simple gaussian shape for the resonance line. However, the position, width and shape of a resonance may be affected by the presence of other resonances or a continuous background because of the configuration interactions.

The central issue of this letter is to point out that Fano-resonances can be expected to be of particular importance for the continuum dynamics of exotic nuclei. In extremely neutron-rich nuclei they may occur close to threshold and would then affect directly the astrophysical capture reactions. In section 2, we present a simple but realistic model for BSEC in exotic nuclei. The model extends the previous work of refs. [14, 15] by incorporating information from Hartree-Fock-Bogoliubov (HFB) and quasi-particle RPA (QRPA) calculations for potentials, single particle excitation energies and transition probabilities, respectively. In section 3, the model is applied to $^{15}$C and used to explain the recently observed structures above the particle threshold. The paper closes in section 4 with a summary and the conclusions.

## 2. Theoretical Approach and Numerical Methods

A theoretical model for the study of resonances and their line shapes in the low-energy continuum is the Quasiparticle-Core Coupling (QPC) model [17]. The QPC Hamiltonian has the form:

$$H = \begin{pmatrix} H_{11} & V_{13} \\ V_{31} & H_{33} \end{pmatrix} \qquad (1)$$

where $H_{11}$ represents the Hamiltonian operator acting on the one-quasiparticle (1-QP) components, describing the single particle motion with respect to an inert core and defining the basis of the single particle states. $H_{33}$ acts on the 3-QP components, chosen as 2-QP QRPA core excitations to which the 1-QP states are coupled. The residual interaction $V_{13}$ couples the 1-QP and 3-QP states. The eigenstates $|\varphi_J\rangle$ of the QPC Hamiltonian H, i.e., $(H - E)|\varphi_J\rangle = 0$, will be therefore a superposition of two components: the 1-QP states $|n_J\rangle$ and the core-excited 3-QP states $|(j'J_C)_J\rangle$, respectively,

$$|\varphi_J\rangle = \sum_n z_n(E)|n_J\rangle + \int d\varepsilon\, z_\varepsilon(E)|\varepsilon_J\rangle + \sum_{j'J_C} z_{j'J_C}(E)|(j'J_C)_J\rangle \qquad (2)$$

where the total spin and parity must be conserved: $\vec{J} = \vec{j}' + \vec{J}_C$; $\pi_J = \pi_{j'} \cdot \pi_{J_C}$, with $\vec{j}'$ and $\vec{J}_C$ indicating the spin of the 1-QP neutron state and the 2-QP core-excited one, respectively. The first term defines the leading particle which, for a $0^+$-core, fixes the total spin J and the parity of the whole configuration. It includes a superposition of radial wave functions from the discrete part of the spectrum with nodes *n* because in an interacting system the radial quantum numbers *n* are no longer conserved. The continuum part of the single particle, i.e. the 1-QP, spectrum is taken into account by the second term. The core excitations in the third term are described by QRPA states with energy $E_C$.

In fact, since here we consider states above the particle threshold the discrete 1-QP components are always off their energy shell. Moreover, without loss of generality we can refrain from an expansion into the basis of mean-field wave functions. Instead, we project the Schrödinger equation onto the 1-QP and 3-QP components and integrate the whole system numerically. After integrating out the core degrees of freedom, the following set of N coupled equations is obtained:

$$\left(h_j^{(1)} - \varepsilon_1\right)\phi_j + \sum_{j'J_C}\langle 0|V_{13}|J_C\rangle \phi_{j'J_c} = 0 \qquad \text{channel 1} \qquad (3a)$$

$$\left(h_{j'J_C}^{(i)} - \varepsilon_i\right)\phi_{j'J_c} + \sum_n \langle J_C|V_{13}|0\rangle \phi_j = 0 \qquad \text{channel i = 2, ..., N.} \qquad (3b)$$

Here, $\phi_j$ and $\phi_{j'J_c}$ denote single particle wave functions with respect to the core ground state and the core-excited states, respectively, determined by the corresponding effective single particle Hamiltonians $h_j^{(1)}$ and $h_{j'J_C}^{(i)}$. The single particle energies are related by $\varepsilon_i = \varepsilon_1 - E_C^{(i)}$, where $\varepsilon_1 = E - E_0$ is the ground state single particle (HFB) energy and $E_C^{(i)}$ is the excitation energy of the core state in channel i. Negative $\varepsilon_1$ values correspond to the bound single particle states, while the positive ones to the 1-QP continuum. The BSEC include bound core-excited states which are represented by negative $\varepsilon_i$ values although their total energy is positive, i.e., they are immersed into the single particle continuum. A consequence of $\varepsilon_i < 0$ is that for $r \to \infty$ the core components $\phi_{j'J_c} \to 0$ are exponentially decaying. Hence, for BSEC-coupling the flux of the elastic component $\phi_j$ is asymptotically conserved, which is different from other channel coupling phenomena. We also emphasize that this formulation fully accounts for channel coupling and mixtures between the discrete and continuum parts of the spectrum, thus avoiding cut-off effects which necessarily occur by using an expansion into a finite set of basis functions.

There are different techniques in use to solve the above coupled channels problem. We choose to work in coordinate space, leading directly to the radial wave functions and the scattering phase shifts as a function of the energy. For a coupled channels problem with N channels there exist N independent elementary solutions $\chi_n$, n = 1, …, N. They are obtained by projecting out the spin-angle degrees of freedom, leading to a set of N coupled equations for the radial wave functions $u_{i,\ell}(r)$

$$\left[\frac{d^2}{dr^2} - \frac{\ell_1(\ell_1+1)}{r^2} + K_1^2\right] u_{1,\ell_1}(r) - \sum_{i=2}^{N} W_i \, u_{i,\ell_i}(r) = 0 \qquad \text{channel 1} \qquad (4a)$$

$$\left[\frac{d^2}{dr^2} - \frac{\ell_i(\ell_i+1)}{r^2} + K_i^2\right] u_{i,\ell_i}(r) - W_i \, u_{1,\ell_1}(r) = 0 \qquad \text{channel } i = 2, …, N \qquad (4b)$$

which are solved numerically. The local wave numbers are given by $K_i^2 = 2m(\varepsilon_i - U_i)/\hbar^2$, approaching at $r \gg R_A$, $R_A$ the nuclear radius, the asymptotic values $k_i^2 = 2m\varepsilon_i/\hbar^2$, where m is the reduced mass in the neutron-core system. The channels are coupled by $W_i = 2m F_{J_C}^{(i)}/\hbar^2$, where $F_{J_C}^{(i)}$ is the reduced transition form factor. Couplings between the closed channels are omitted.

Once we have determined the set of fundamental solutions $\chi$, they are used as basis to construct the physical solutions by superposition up to the matching radius $R_M$:

$$u_{i,\ell_i}(r) = \sum_{m=1}^{N} b_m \, \chi_{i\,m}(r) \qquad\qquad i = 1, …, N \qquad (5)$$

The coefficients $b_m$ are determined by imposing the boundary conditions that both $u_{i,\ell}(r)$ and $du_{i,\ell}(r)/dr$ coincide at $r = R_M$ with the asymptotic forms. In our case this means to form a superposition complying with asymptotically incoming and outgoing waves present only in the ground state elastic channel and all other channels are asymptotically closed. Hence, at radii $r \geq R_M \gg R_A$ the radial wave functions of the neutron must obey:

$$u_{i,\ell_i}(r) = j_{\ell_1}(k_1 r)\,\delta_{1i} + C_{1i}\, h^{(+)}_{\ell_i}(k_i r) \qquad (6)$$

Spherical Bessel and Hankel functions are denoted by $j_\ell(x)$ and $h^{(+)}_\ell(x)$, respectively. Above, for i ≠

1, $k_i \equiv i\kappa_I$, is a purely imaginary quantity, implying exponentially decaying wave functions in the corresponding channels. In particular, we are interested in the scattering matrix elements in the open channel because they determine the partial wave elastic cross sections $\sigma_{ii} = \sigma_i(k_i)$ (here only for i = 1). For a particle with spin *s* they are given by:

$$\sigma_i(k_i) = \frac{4\pi}{k_i^2} \frac{2j+1}{2s+1} |C_{ii}(k)|^2 \qquad (7)$$

## 3. Bound States Embedded in the Continuum in $^{15}$C

Calculations have been performed for continuum states in $^{15}C = n + {}^{14}C$ because the aforementioned charge exchange data indicate sharp spectral structures of unusual line shape at low excitation energies. Here, we present results of exploratory calculations for d-wave elastic scattering states which we find to be especially sensitive on the core coupling effects. The reason is that the HFB potential predicts for the $d_{3/2}$ partial wave an elastic scattering potential resonance at about $E_R = 3.6$ MeV above the threshold of a width $\Gamma \sim 2.1$ MeV. The existence of a single particle resonance is an important prerequisite for strong continuum DCP effects in that particular partial wave. The reason is quite obvious because the coupling to the more complex core-excited states relies on a good overlap of the single particle scattering wave with the nuclear interior. This condition is given in the vicinity of a resonance.

In the calculations we have used HFB potentials as reference. The parameters were slightly adjusted such that the experimental (if available) or HFB single particle energies and root-mean-square radii of the single neutron states in the $^{14}$C core system are reproduced. Five channels are included into the calculations for $^{15}$C: the single particle continuum and the four core-excited states in $^{14}$C at $E_C$ ($J^\pi$) = 6.094 ($1^-$), 6.728 ($3^-$), 7.012 ($2^+$), 8.317 ($2^+$) MeV. The particle-core interactions $W_i$ of eqs. (4) are described by an average value $F_0 \sim V_{13}$ which we treat as a free parameter, and state dependent core transition amplitudes $\beta_{J_C}^{(i)}$. They are identified to be the collective deformation amplitudes, e.g., observed in the $^{14}C(\alpha,\alpha')$ reactions [18].

The n + $^{14}$C $d_{5/2}$ partial wave elastic cross section $\sigma_{11}$ is displayed in Fig. 1a for excitation energies $E_x = 0, \ldots, 12$ MeV in $^{15}$C. Several narrow, long-living resonance structures above the particle threshold appear in the energy region between 6 and 10 MeV in the $d_{5/2}$-wave cross section once the particle-core interaction is turned on, while no resonances are observed in the s- and p-waves. Their asymmetric line shape results from the bound-continuum coupling, as typical for Fano-resonances.

An important prediction of Fano's theory, which later was confirmed also for the nuclear case [13, 14, 15], is the direct relation between the coupling matrix elements and the width of a Fano-resonance. This dependence is also found here, seen when varying the interaction strength $F_0$ to our will. For small $F_0$ (1-2 MeV) the coupling is so weak that only small fluctuations appear on a continuous background. For intermediate $F_0$ the resonances become more intense and narrow until, starting from $F_0 \sim 7$ MeV, their width is increasing progressively until individual states are no longer resolved. Hence, measurements of resonance line shapes will provide information on residual interactions in exotic nuclei.

The results shown in Fig. 1a, where well resolved states and visible interferences can be observed, are obtained with $F_0 = 5$ MeV, a value which seems to be a rather good compromise. From an experimental point of view, these interferences may be hard to observe if the resonance is not well isolated because of insufficient experimental resolution or additional background contributions. However, the asymmetry of the 8.5 MeV peak observed in the $^{15}$C spectra of refs. [8, 9] is a convincing signature for a BSEC excitation, constituting experimental evidence for such core polarization effects.

The theoretical cross section (Fig. 1a) is compared to the $^{15}$C experimental spectrum taken at $\theta_{lab} = 14°$ [8, 9], shown in Fig. 1b in the energy region of interest. This comparison is on a qualitative level since the present calculations do not account for the reaction dynamics of the producing charge exchange reaction. However, we do describe the rescattering of the neutron after production on the residual target nucleus which is responsible for the structures seen in the measured spectrum. With our standard parameter set four theoretical levels are found at $E_x$ = 6.67, 7.36, 7.70, 8.92 MeV of widths $\Gamma$ = 66, 80, 141, 85 keV, respectively. These widths are estimate from Gaussian fits to the peaks, even though we have to keep in mind that they are asymmetric. The experimental resonances are observed at $E_x = (6.77 \pm 0.06, 7.30 \pm 0.06, 8.50 \pm 0.06)$ MeV and Gaussian fits lead to the estimated widths $\Gamma \leq$ 160, 70, 140 keV, respectively [8, 9]. Assuming an additional average energy shift of 400 keV, the theoretical peaks become $E_x$ = 6.27, 6.96, 7.30 and 8.52 MeV. These values are surprisingly close to the experimental ones.

### 4. Conclusions

Fano-resonances in β-unstable neutron-rich nuclei have been studied theoretically in a coupled channels model. The approach described here, also extending the QPC model into the continuum, explains the resonance line shape as an effect of the interference between an open 1-QP and closed 3-QP channels. The calculations reproduce in a qualitative way the narrow states observed experimentally in the $^{15}$C* = n + $^{14}$C continuum.

Similar calculations are planned for the more exotic carbon isotopes, $^{17}$C and $^{19}$C, in order to investigate systematically the BSEC phenomenon and the interference effects in neutron-rich nuclei far off stability. The influence of different open channels will be investigated in nuclei where they have comparable thresholds as, e.g., $^{12}$C.

The study of the low-lying resonances in light exotic nuclei is of astrophysical importance also. An interesting aspect, indeed, is that the BSEC phenomena lead to a change of the level density close to the continuum threshold. Hence, neutron capture processes will be affected which, in turn, might influence astrophysical reaction rates and nuclear life times. The increase of the level density beyond the typically assumed density of states in a potential will also affect the thermodynamical properties of exotic nuclei.

### Acknowledgements

This work is supported in part by the MCTS program of the European Community, contract No. HPMT-CT-200100223, and GSI Darmstadt.

**References**

[1] Heyde K., Basic Ideas and Concepts in Nuclear Physics, edited by Brewer D. F., IOP Publishing, Bristol and Philadelphia (1999), Ch. 14.

[2] Tanihata I., Nucl. Phys. A 520 (1990) 411; Tanihata I., Nucl. Phys. A 522 (1991) 275c.

[3] Riisager K., Jensen A. S. and Moller P., Nucl. Phys. A 548 (1992) 393.

[4] Hansen P. G., Jensen A. S. and Jonson B., Annu. Rev. Nucl. Part. Sci. 45 (1995) 591.

[5] Kobayashi T., Phys. Rev. Lett. 60 (1988) 2599.

[6] Lenske H., J. Phys. G: Nucl. Part. Phys. 24 (1998) 1429.

[7] Cappuzzello F., Cunsolo A., Fortier S., Foti A., Khaled M., Laurent H., Lenske H., Maison J. M., Melita A. L., Nociforo C., Rosier L., Stephan C., Tassan-Got L., Winfield J. S. and Wolter H. H., Phys. Lett. B 516 (2001) 21.

[8] Orrigo S. E. A., Allia M. C., Beaumel D., Cappuzzello F., Cunsolo A., Fortier S., Foti A., Lazzaro A., Lenske H., Nociforo C. and Winfield J. S., Proceeding "10th International Conference on Nuclear Reaction Mechanisms", Villa Monastero, Varenna, Italy, 9-13/06/2003, edited by E. Gadioli, Ricerca Scientifica ed Educazione Permanente (2003), Supplemento n.122, p. 147, available at http://lxmi.mi.infn.it/~gadioli/proceedings/orrigo.doc; Orrigo S. E. A., Ph.D. Thesis, University of Catania (2004) available at http://eprints.ct.infn.it/29/.

[9] Cappuzzello F., Orrigo S. E. A., Cunsolo A., Lenske H., Allia M. C., Beaumel D., Fortier S., Foti A., Lazzaro A., Nociforo C. and Winfield J. S., Europhys. Lett. 65 (2004) 766.

[10] Fano U., Phys. Rev. 124 (1961) 1866.

[11] Glutsch S., Physical Review B 66 (2002) 075310 and refs. therein.

[12] Ligterink N. E., PiN Newslett. 16 (2002) 400, e-archive: nucl-th/0203054 (2002).

[13] Mahaux C. and Weidenmüller H. A., Shell Model Approach to Nuclear Reactions, North-Holland, Amsterdam (1969).

[14] Baur G. and Lenske H., Nucl. Phys. A 282 (1977) 201.

[15] Fuchs H., Nolen J. A., Wagner G. J., Lenske H. and Baur G., Nucl. Phys. A 343 (1980) 133.

[16] Cortina-Gil D., Baumann T., Geissel H., Lenske H., Sümmerer K., Axelsson L., Bergmann U., Borge M. J. G., Fraile L. M., Hellström M., Ivanov M., Iwasa N., Janik R., Jonson B., Markenroth K.,

Münzenberg G., Nickel F., Nilsson T., Ozawa A., Riisager K., Schrieder G., Schwab W., Simon H., Scheidenberger C., Sitar B., Suzuki T. and Winkler M., Eur. Phys. J. A 10 (2001) 49.

[17] Lenske H., Keil C. M. and Tsoneva N., J. Progr. Part. Nucl. Phys. 53 (2004) 153; Nociforo, C. and Lenske H., in preparation.

[18] Ajzenberg-Selove F., Nucl. Phys. A 523 (1991) 1 and refs. therein.

**Figure 1:** a) Calculated elastic cross section $\sigma_{11}$ for the $d_{5/2}$-wave plotted with respect to the $^{15}$C excitation energy. Core states included in the calculations: $E_C$ ($J^\pi$) = 6.094 (1$^-$), 6.728 (3$^-$), 7.012 (2$^+$), 8.317 (2$^+$) MeV. b) Excitation energy spectrum for the $^{15}$N($^7$Li,$^7$Be)$^{15}$C reaction at $E_{inc}$ = 55 MeV and $\theta_{lab}$ = 14° [8, 9]. Peaks marked with an asterisk correspond to the mutual excitation of $^7$Be at 0.429 MeV. The continuous line is the sum of the background (the $^{15}$N($^7$Li,n$^7$Be)$^{14}$C 3-body phase space) and the Gaussians from the fit to the peaks.

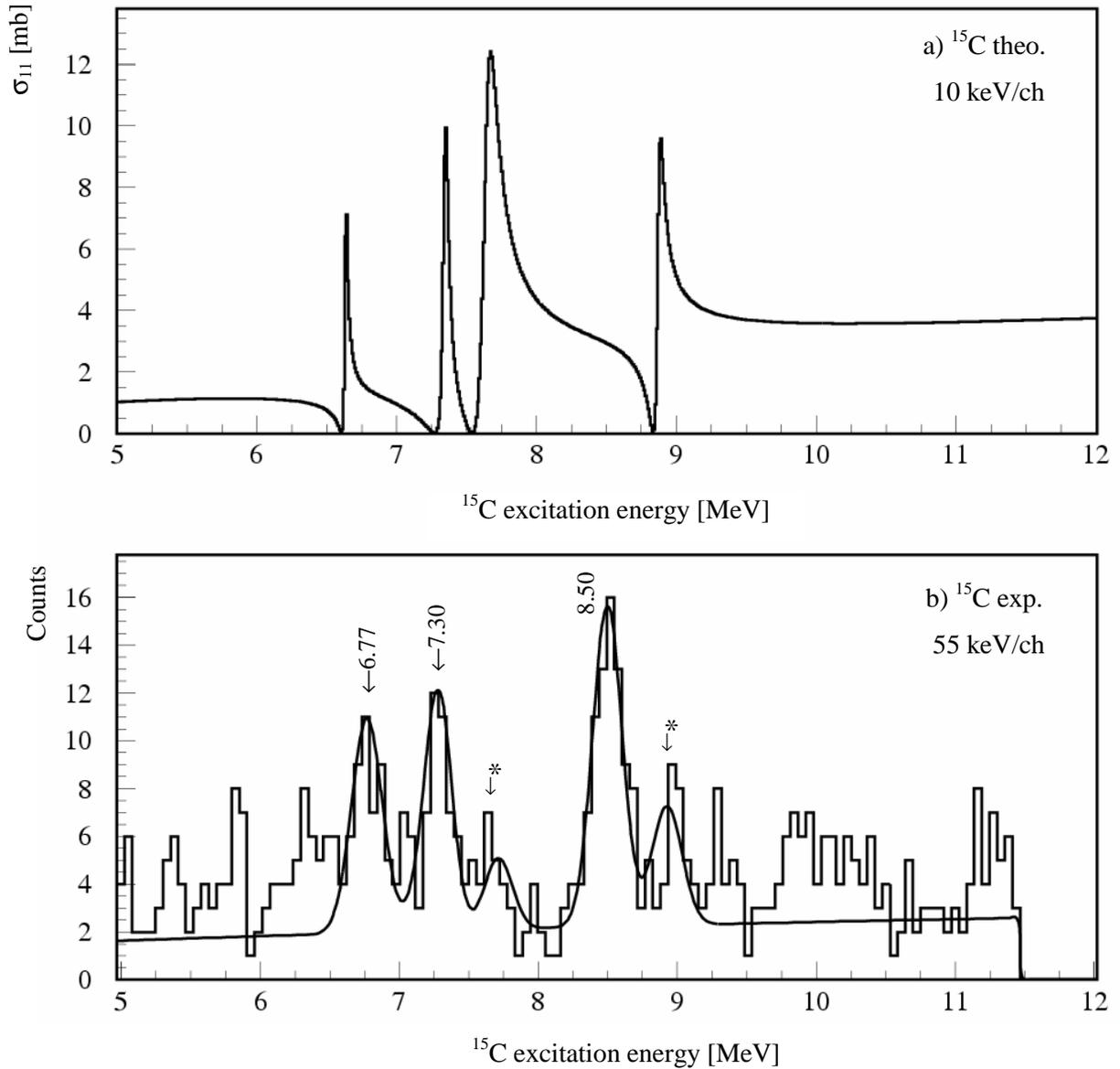